# Why Blind-Variation Selective-Retention is an Inappropriate Explanatory Framework for Creativity

Comment on "Creative thought as Blind-variation and Selective-retention: Combinatorial Models of Exceptional Creativity"
by Dean Keith Simonton

Liane Gabora
University of British Columbia

Abstract

Simonton is attempting to salvage the Blind Variation Selective Retention theory of creativity (often referred to as the Darwinian theory of creativity) by dissociating it from Darwinism. This is a necessary move for complex reasons outlined in detail elsewhere. However, whether or not one calls BVSR a Darwinian theory, it is still a variation-and-selection theory. Variation-and-selection was put forward to solve a certain kind of paradox, that of how biological change accumulates (that is, over generations, species become more adapted to their environment) despite being discarded at the end of each generation (that is, parents don't transmit to offspring knowledge or bodily changes acquired during their lifetimes, e.g., you don't inherit your mother's ear piercings). This paradox does not exist with respect to creative thought. There is no discarding of acquired change when ideas are transmitted amongst individuals; we share with others modified versions of the ideas we were exposed to on a regular basis.

Campbell was an important philosopher and an original thinker, and the endeavor to revitalize one of his more controversial ideas, the idea that creative thought proceeds through blind variation and selective retention (BVSR) is admirably ambitious. Simonton's [7] book on the topic won awards, and in the Cambridge Handbook of Creativity, BVSR is described as: "arguably the most ambitious account of Big-C Creativity. It has contributed a very rich repository of results and ideas and numerous specific quantitative predictions, and many (but not all) of its claims boast substantial support" [6]. However, this paper, though interesting and well written, has serious problems.

Simonton appears to be attempting to salvage BVSR by dissociating it from Darwinism. This is a necessary move for complex reasons that have been outlined in detail elsewhere (Gabora [3,4]), albeit a significant departure from previous writings by Campbell and Simonton. The problem is: whether or not you call BVSR a Darwinian theory, it is still a variation-and-selection theory, and variation-and-selection was put forward to solve a certain kind of paradox, that of how biological change accumulates — that is, over generations, species become more adapted to their environment—despite being discarded at the end of each generation—that is, parents don't transmit to offspring knowledge or bodily changes acquired during their lifetimes, e.g. you don't inherit your mother's ear piercings. This paradox does not exist with respect to creative thought. There is no discarding of acquired change when ideas are transmitted amongst individuals; we share with others modified versions of the ideas we were exposed to on a regular basis (see [4]).

Variation-and-selection is useful for explaining adaptive change when there is no mechanism for preferentially generating high quality variants. Lots of possibilities are randomly generated (the variation-generating phase), and the weeding out occurs later, at the selection phase (the variation-discarding phase). But with respect to creativity, there is no need for this kind of explanatory framework because the generation process is biased toward ideas that stand a good chance of being valuable. (Indeed the author mentions several of the mechanisms responsible for this in this paper: expertise, remote associations, etc.) The bottom line is that it is not clear what a variation–selection framework buys you with respect to creativity beyond more conventional two-stage theories such as Finke, Ward, and Smith's [2] Geneplore model, which proposes that creativity consists of a generative stage followed by an exploratory phase.

Stripped of the explanatory strength of the Darwinian framework, what remains of BVSR is the hypothesis that creativity involves 'blind variations' and selection, but application of the concepts 'blindness', 'variation', and 'selection' to creativity are all questionable. It goes without saying that a creator does not know at the outset exactly how a creative task will be accomplished, so a theory of creativity that distinguishes itself on the basis of claims about 'blindness' seems less than useful. The appropriateness of the term 'variation' in biology stems from the fact that it is possible to accurately and objectively measure the relatedness of entities. Whether or not two organisms share a common ancestor is clear-cut; they either are or are not descendents of a particular individual. But with respect to creativity, where concepts and ideas are constantly combined, contrasted, and reinterpreted in light on each other, there is no basis upon which to delineate those ideas that are or are not variants of a given idea. Simonton's application of the notion of selection is also problematic. The generation of one idea affects the conception of the task, and thus the criteria by which the next is judged. Therefore, successively generated ideas cannot be treated as members of a generation, and are not selected amongst.

Creativity is perhaps what more than anything else separates human cognition from that of other species, and as such the development of theories of creativity both presents considerable challenges and holds considerable promise. BVSR is arguably one of the best-known efforts to explain creativity, and I imagine many will be interested to see its latest incarnation. However, it is not clear that this version of the theory successfully overcomes problems that were present in previous versions.